# Technical Brief

# Improved mass spectrometry compatibility is afforded by ammoniacal silver staining

Mireille Chevallet[1], Hélène Diemer[2], Sylvie Luche[1], Alain van Dorsselaer[2], Thierry Rabilloud[1] and Emmanuelle Leize-Wagner[2,+].

[1] CEA- Laboratoire d'Immunochimie, DRDC/ICH, INSERM U 548
CEA-Grenoble, 17 rue des martyrs, F-38054 GRENOBLE CEDEX 9, France

[2] Laboratoire de Spectrométrie de Masse Bio-Organique, UMR CNRS 7512, ECPM, 25 rue Becquerel, 67087 STRASBOURG Cedex2, France

+ present address : Institut de Sciences et ingéniérie Supramoléculaire, UMR CNRS 7006, Université Louis Pasteur, 8 rue Gaspard Monge, 67083 STRASBOURG CEDEX, France

Correspondence :
Thierry Rabilloud, DRDC/ICH, INSERM U 548
CEA-Grenoble, 17 rue des martyrs,
F-38054 GRENOBLE CEDEX 9
Tel (33)-4-38-78-32-12
Fax (33)-4-38-78-98-03
e-mail: Thierry.Rabilloud@ cea.fr

Abstract
Sequence coverage in mass spectrometry analysis of protein digestion-derived peptides is a key issue for detailed characterization of proteins or identification at low quantities. In gel-based proteomics studies, the sequence coverage greatly depends on the protein detection method. It is shown here that ammoniacal silver detection methods offer improved sequence coverage over standard silver nitrate methods, while keeping the high sensitivity of silver staining

With the development of 2D-PAGE-based proteomics, another burden is placed on the detection methods used for protein detection on 2D-gels. Besides the classical requirements of linearity, sensitivity and homogeneity from one protein to another, detection methods must now take into account another aspect, namely their compatibility with mass spectrometry. This compatibility is evidenced by two different and complementary aspects, which are i) the absence of adducts and artefactual modifications on the peptides obtained after protease digestion of a protein detected and digested in –gel and ii) the quantitative yield of peptides recovered after digestion and analyzed by the mass spectrometer. While this quantitative yield is not very important per se, it is however a crucial parameter as it strongly influences the signal to noise ratio of the mass spectrum, and thus the number of peptides that can be detected from a given protein input, especially at low protein amounts. This influences in turn the sequence coverage, and thus the detail of the analysis provided by the mass spectrometer.

Several detection methods are widely used for protein visualization after 2D gel electrophoresis. The most popular one is probably colloidal Coomassie blue staining [1]. Although this method is not very sensitive, it affords a very good compatibility with mass spectrometry. It also has gained popularity in the early days of proteomics, when its sensitivity matched almost exactly the needs in protein amounts required for a decent analysis by mass spectrometry. With the improvement of the mass spectrometers, associated robotics etc…, this is no longer the case, and detection by colloidal Commassie either requires useless consumption of sample or leads to low numbers of analyzable proteins due to undetection. Fluorescent detection methods offer an interesting alternative, and metal chelate-based methods [2, 3] have become increasingly popular. In their most recent versions [4], these methods offer an interesting sensitivity allied to a very good compatibility with mass spectrometry analysis (see for example [3]). These methods are however not easy to use in small proteomics laboratories. Besides the expensive laser scanners or CCD cameras required for quantitative analysis, spot excision must be performed under UV light, which induces in turn collateral problems such as safety problems and photobleaching problems, both increasing when large number of spots are to be excised in comprehensive proteomics studies.

Last but not least, silver staining still offers the maximal sensitivity, and all the ancillary advantages associated on light absorption-based methods, such as easy visualization and quantitation (although the linear dynamic range of silver staining is not very good), and easy spot excision. However, silver staining is plagued by a rather poor compatibility with mass spectrometry. This has been exemplified in numerous papers (e.g. in [5] and [6]) and is only partially alleviated by destaining procedures [7] after silver staining and prior to protein digestion. The interference between silver staining and mass spectrometry has been investigated in detail [8], and this has evidenced the crucial role of the formaldehyde used in all silver staining methods as the silver-reducing agent. This puts forward an interference mechanism in which protein crosslinking by formaldehyde in alkaline media could be the major phenomenon taking place. If this mechanism is true, ammoniacal silver methods [9] should provide an increased compatibility with mass spectrometry, as the development step takes place in an acidic medium and is made in the presence of the excess ammonia carried over with silver in the silvering agent, this ammonia acting as a scavenger for formaldehyde.

We therefore tested silver nitrate, glutaraldehyde-free silver ammonia and RuBPS as detection methods after 2D PAGE. As shown on figure 1, both silver nitrate and silver ammonia give similar detection thresholds (see also [9]) which were better than those obtained with RuBPS. We then selected proteins of various abundances, hydrophobicities, Mw and pI for mass spectrometry analysis, which was carried out by Peptide Mass Fingerprinting with a MALDI instrument according to classical methods [3]. For silver-stained gels, with silver nitrate or silver ammonia, the excised spots were destained by ferricyanide-thiosulfate [7] as soon as possible after silver staining, as this was shown to improve the mass spectrometry results [8]. The results are summarized on table 2. They clearly show that protein behaviours are variable but show however a clear trend that can be summarized as follows. Silver-stained spots give signals that are at best equal to those obtained with RuBPS (e.g. Band 3) but that can also be inferior. In the latter case, silver-ammonia and the Shevchenko methods gives results that are superior or equal to those obtained with silver nitrate (e.g. PGM, Ran, endoplasmin, hsp90). In trying to go further and understand why some peptides are missing in silver-stained gels compared to RuBPS-stained gels, we could not find obvious trends (e.g. selective loss of lysine-containing peptides compared to arginine-containing peptides, influence of content in nucleophilic amino acids such as Ser, Cys, Thr or Tyr etc…) so that the exact mechanism of peptide losses is still unknown.

On a more practical basis, it can be concluded that silver ammonia is an interesting choice over silver nitrate as a compromise between sensitivity, linearity, absence of background, ease of use, cost (including hardware) and mass spectrometry compatibility. However, it must be kept in mind that ammoniacal silver is not compatible with all electrophoresis systems. Although it is fully compatible with the popular Tris-glycine system and the more recent Tris taurine system [13], it is not compatible with the Tris-tricine systems [14, 15]. It can also be rendered compatible with supported gels [16], but in all cases optimal performance is obtained when thiosulfate is included in the gel upon polymerization [17], which precludes the use of ready-made gels. However, in this latter case, substantial background reduction can be obtained by a short rinse in a dilute solution of thiosulfate just before development [18]. Last but not least, optimal performance strongly depends on the silver-ammonia ratio [19], which controls both the sensitivity and the tendency to background staining. This parameter is to be considered as critical in long-term experiments, due to the short shelf-life of ammonia solutions. A good procedure for controlling this parameter is to use ready-to-use, titrated ammonium hydroxide solutions.

When these conditions are met, ammoniacal silver is clearly the best silver-based detection methods for proteomics studies. While the sequence coverage is often inferior to the ones obtained with the Shevchenko methods, the background due to carrier ampholytes [20] and chromatic effects encountered with the latter method preclude its use as a quantitative method. This strongly decreases the interest of the method, as one of the main interests of the silver-ammonia method is the ability to use the same experiment for the quantitative image analysis and mass spectrometry analysis of spots at a much inferior cost than fluorescent staining.

Legends to figures

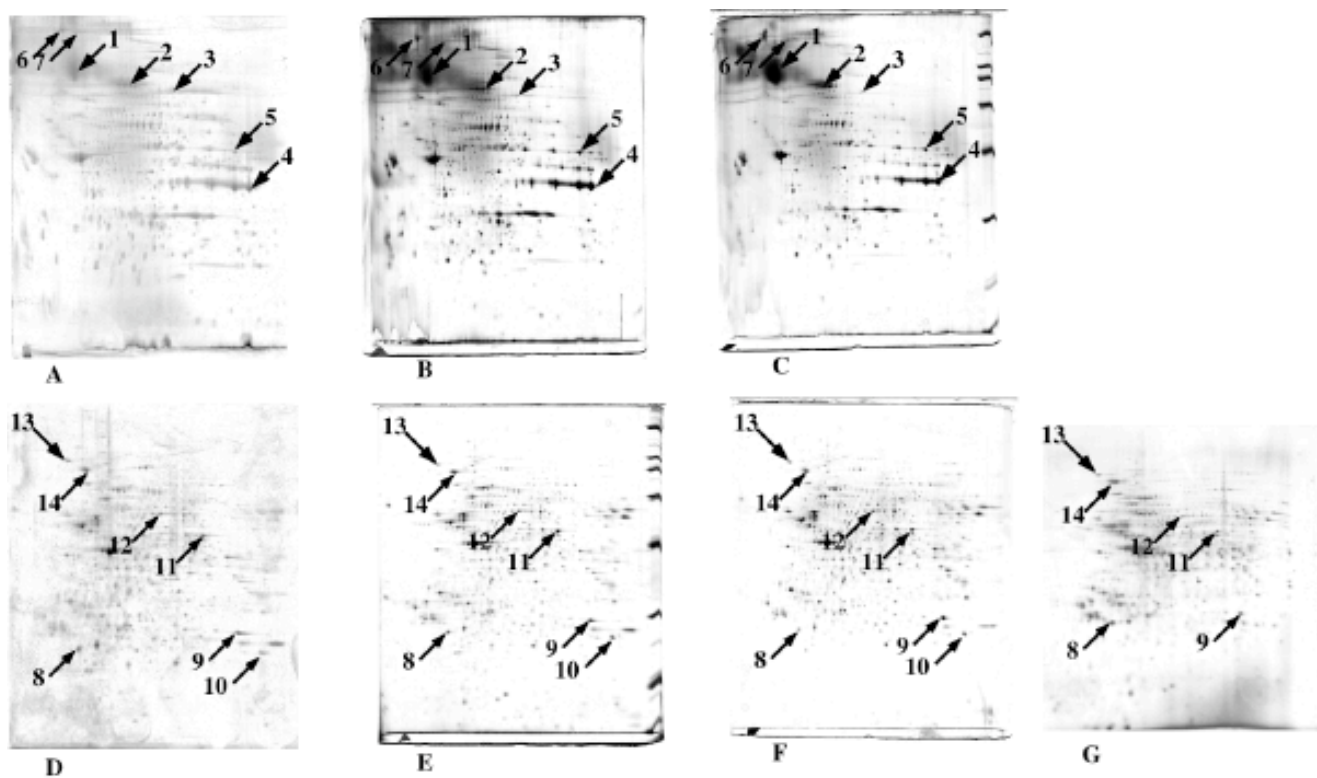

**Figure 1:** Comparative analysis of proteins by two-dimensional electrophoresis.
2D gel separation conditions: IEF range, pH 4-8 or 4-10 70 kVh, Mw separation on a 10% gel operated in taurine system [13]. Protein load: 100 μg, protein extraction in 7M urea, 2M thiourea and detergent. Top row (panels A to C): human erythrocyte membrane proteins (detergent 2% ASB 14, IEF 4-10). Bottom row (panels D to G): J774 mouse macrophage cell line (detergent 4% CHAPS, IEF 4-8). Left column (panels A and D): detection by RuBPS. Center column (panels B and E): detection by silver nitrate [10]. Right column (panels C and F): detection with ammoniacal silver [11]. G. detection with the Shevchenko fast silver staining method [13]. The latter method gives a strong background In the low molecular weight area of the gel due to carrier ampholytes. Furthermore, it gives strong color effects in spots, which precludes any quantitative analysis.
The numbered spots indicated by arrows have been excised for further analysis by mass spectrometry

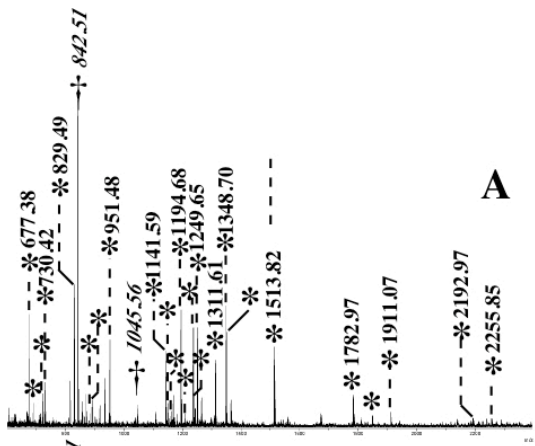
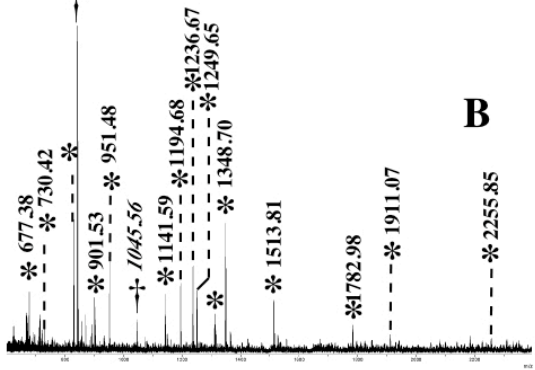
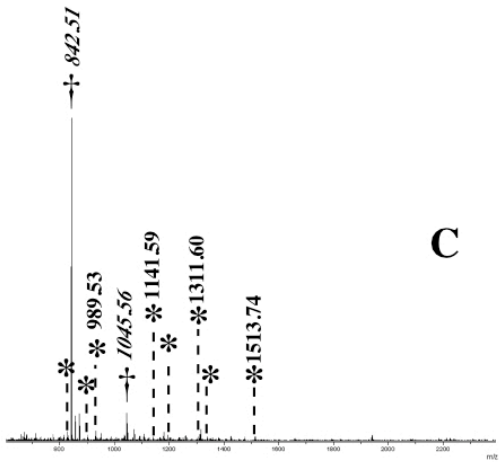
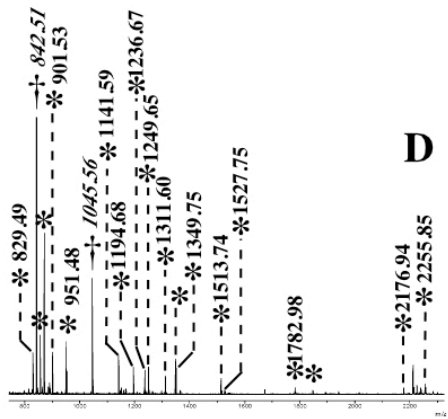

**Figure 2:** Comparison of peptide mass fingerprinting spectra.

The MALDI mass fingerprinting spectra obtained for the hsp90 protein (SwissProt P08238) are shown Peaks are automatically collected with a signal to noise ratio above 4 and a peak quality index greater than 30.

A) spectrum obtained after ruthenium fluorescent staining [4]

B) spectrum obtained after ammoniacal silver staining [11]

C) spectrum obtained after standard silver nitrate staining [10]

D) spectrum obtained after silver staining as described by Shevchenko et al. [12]

*: peptide masses matching with the predicted ones

†: trypsin autolysis peptides

List of matching masses [the signal to noise ratio is given between parentheses for selected peaks (in bold) ]:

C) Silver nitrate staining: [**951.47** (19.4); **1141.56** (11) **1513.75** (12)]; 829.53; 901.52; 1194.64; 1311.57; 1348.66;

B) ammoniacal silver: same as above [**951.47** (40.5) ; **1141.56** (29.2) **1513.75** (25.9)] plus: : 677.39; 1151.56; 1236.64; 1249.62; 1782.89; 1910.98; 2255.85

A) RuBPS: same as B [**951.47** (83); **1141.56** (73); **1513.75** (96.5); ] plus: 689.38; 722.48; 730.44; 886.55; 891.43; 1160.58; 1208.63; 1242.70; 1308.67; 1349.75 , 1808.95; 1847.80; 2192.98

D) same as C [**951.47** (141.2); **1141.56** (111.3); **1513.75** (61.9) ] plus: 886.55; 891.43, 1151.56, 1160.58, 1208.63; 1236.64; 1249.62, 1349.75, 1376.65, 1527.75, 1782.89; 1847.80, 2176.94, 2255.85

Note the signal reduction in C compared to the other conditions, at equal protein load.

Table 1: flowchart for silver ammonia staining

| Step | solution | Time |
|---|---|---|
| Fixation | Ethanol 30% (v/v) Acetic acid 10% (v/v) 2,7 Naphthalene disulfonic acid 0.05% | overnight |
| Rinse | water | 6x 10 minutes |
| Silvering | Silver nitrate 24 mM, NaOH 15mM, ammonium hydroxide 75mM | 20-30 minutes |
| Rinse | water | 3 x 5 minutes |
| development | 37% formaldehyde: 1 ml/l, citric acid 80 mg/l | 5-10 minutes |
| Stop | Ethanolamine 5 ml/l, acetic acid 20 ml/l | 30 minutes |

Polymerization initiator for inclusion of thiosulfate in gels: use 0.7µl TEMED, 6µl of 10% sodium thiosulfate solution and 7µl of 10% ammonium persulfate solution (to be added last) per ml of gel solution

Table 2: mass spectrometry analysis of various protein spots

| Spot Number | identification | Accession number | Detection method | Sequence coverage | Number of matched peptides |
|---|---|---|---|---|---|
| 1 | Band 3 Anion Transport Protein | P02730 | RuBPS | 32% | 26 |
|   | Theoret Mw/pI: 101727 / 5,08 |   | Ag | 17 % | 16 |
|   |   |   | Ag-NH3 | 30% | 30 |
| 2 | Protein 4.1 | P11171 | RuBPS | 34% | 33 |
|   | Theoret Mw/pI: 96957 / 5,45 |   | Ag | 38% | 32 |
|   |   |   | Ag-NH3 | 39% | 38 |
| 3 | Protein 4.2 | P16542 | RuBPS | 21% | 13 |
|   | Theoret Mw/pI: 76794 / 8.27 |   | Ag | NA | 0 |
|   |   |   | Ag-NH3 | NA | 0 |
| 4 | GAPDH | P04406 | RuBPS | 57% | 24 |
|   | Theoret Mw/pI: 35899 / 8,58 |   | Ag | 69% | 26 |
|   |   |   | Ag-NH3 | 68% | 27 |
| 5 | dematin | Q08495 | RuBPS | 49% | 21 |
|   | Theoret Mw/pI: 45486 / 8,94 |   | Ag | 33% | 31 |
|   |   |   | Ag-NH3 | 46% | 21 |
| 6 | Spectrin alpha chain | P02549 | RuBPS | 34% | 89 |
|   | Theoret Mw/pI: 279744 / 4,96 |   | Ag | 30% | 72 |
|   |   |   | Ag-NH3 | 25% | 65 |
| 7 | Spectrin beta chain | P11277 | RuBPS | 12% | 26 |
|   | Theoret Mw/pI: 246170 / 5,13 |   | Ag | NA | 0 |
|   |   |   | Ag-NH3 | NA | 0 |
| 8 | Rho GDI 2 | Q61599 | Ru-BPS | 24% | 5 |
|   | Theoret Mw/pI: 22836 / 4,97 |   | Ag | NA | 0 |
|   |   |   | Ag-NH3 | 21% | 4 |
|   |   |   | Shev | 36% | 7 |
| 9 | PGM | P18669 | Ru-BPS | 64% | 14 |
|   | Theoret Mw/pI: 28655 / 6,75 |   | Ag | 31% | 7 |
|   |   |   | Ag-NH3 | 43% | 10 |
|   |   |   | Shev | 37% | 9 |
| 10 | Ran | P17080 | Ru-BPS | 30% | 7 |
|   | Theoret Mw/pI: 24408 / 7,01 |   | Ag | NA | 0 |
|   |   |   | Ag-NH3 | 24% | 6 |

| | | | Shev | ND | ND |
|---|---|---|---|---|---|
| 11 | Alpha enolase | P17182 | Ru-BPS | 58% | 23 |
| | Theoret Mw/pI: 46980 / 6,36 | | Ag | 22% | 9 |
| | | | Ag-NH3 | 38% | 15 |
| | | | Shev | 36% | 13 |
| 12 | PDI | P17182 | Ru-BPS | 49% | 29 |
| | Theoret Mw/pI: 56586 / 5,98 | | Ag | 19% | 10 |
| | | | Ag-NH3 | 19% | 10 |
| | | | Shev | 33% | 13 |
| 13 | endoplasmin | P08113 | Ru-BPS | 32% | 29 |
| | Theoret Mw/pI: 92418 / 4,74 | | Ag | NA | 0 |
| | | | Ag-NH3 | 23% | 24 |
| | | | Shev | 19% | 18 |
| 14 | Hsp 90 beta | P08238 | Ru-BPS | 35% | 27 |
| | Theoret Mw/pI: 83081 / 4,97 | | Ag | 10% | 8 |
| | | | Ag-NH3 | 21% | 15 |
| | | | Shev | 33% | 25 |

Ru-BPS : ruthenium complex fluorescent staining ; Ag : silver nitrate staining [10]. Ag-NH3 : silver-ammonia staining (table 1) ; Shev : silver nitrate staining according to Shevchenko et al. [12]. NA : not available (no identification). ND : not detected by the method selected